\documentclass[10pt,conference]{IEEEtran}
\IEEEoverridecommandlockouts
\usepackage{cite}
\usepackage{amsmath,amssymb,amsfonts}
\usepackage{algorithmic}
\usepackage{graphicx}
\usepackage{textcomp}
\usepackage{xcolor}
\usepackage{booktabs}
\usepackage{xspace}
\usepackage{tabularx}
\usepackage{siunitx}
\usepackage{url}
\usepackage{balance}
\usepackage{pifont}
\usepackage{hyperref}
\usepackage{cleveref}
\usepackage{makecell}
\usepackage{orcidlink}
\usepackage{framed}
\usepackage{csquotes}
\usepackage{pgfplots}
\pgfplotsset{compat=1.18}

\newcommand{\boxedtext}[1]{\fbox{\scriptsize\bfseries\textsf{#1}}}
\newcommand{\nota}[2]{%
  \boxedtext{#1}{\small$\blacktriangleright$\emph{\textsl{#2}}$\blacktriangleleft$}%
}
\makeatletter
\newcommand{\todo}[1]{%
  {\color{red}\nota{TODO}{#1}}%
  \futurelet\todo@next\todo@checknext
}
\newcommand{\todo@checknext}{%
  \ifx\todo@next\@sptoken
    \let\todo@otherwise\space
  \else
    \let\todo@otherwise\@empty
  \fi
  \@ifnextchar\todo{\par\noindent}{\todo@otherwise}%
}
\makeatother

\def\BibTeX{{\rm B\kern-.05em{\sc i\kern-.025em b}\kern-.08em
    T\kern-.1667em\lower.7ex\hbox{E}\kern-.125emX}}

\def\swebench{SWE-bench\xspace}
\def\opus{Claude Opus 5.5\xspace}
\def\gemini{Gemini 4 Argon\xspace}
\def\astra{GPT-6 Astra\xspace}
\newcommand*\elide{\textup{[\,\dots]}\xspace}
\newenvironment{researchquestions}{%
  \par\begingroup\raggedright
  \setlength{\leftskip}{1.75em}%
  \hyphenpenalty=10000\exhyphenpenalty=10000
  \setlength{\parindent}{0pt}%
  \setlength{\parskip}{3pt}%
}{\par\endgroup}

\begin{document}

\title{Construct Validity of Coding Benchmarks for LLM-based Agentic Coding Tools}

\author{\IEEEauthorblockN{Alexander Berndt \orcidlink{0009-0009-5248-6405}$^{*}$ }
\IEEEauthorblockA{
 \textit{Heidelberg University} \\
alexander.berndt@uni-heidelberg.de
}
\and
\IEEEauthorblockN{Paul Ralph \orcidlink{0000-0002-7411-0857}}
\IEEEauthorblockA{\textit{Dalhousie University} \\
paulralph@dal.ca
}
\and
\IEEEauthorblockN{Thomas Bach \orcidlink{0000-0002-9993-2814}}
\IEEEauthorblockA{\textit{SAP} \\
thomas.bach03@sap.com 
}
\and
\IEEEauthorblockN{Sebastian Baltes \orcidlink{0000-0002-2442-7522}}
\IEEEauthorblockA{\textit{Heidelberg University} \\
sebastian.baltes@uni-heidelberg.de 
}
\thanks{$^*$Also affiliated with SAP.}
}

\maketitle

\begin{abstract}
\emph{Background.} Software developers increasingly adopt LLM-based agentic coding tools to assist them with development tasks. To provide developers with up-to-date information on these tools' effectiveness, prior work introduced benchmarks that enable comparisons across tools. Despite the widespread adoption of such benchmarks, however, recent work has shown that they suffer from issues threatening their validity.
\emph{Aims.} In this work, we aim to complement prior work by analyzing the construct validity of agentic coding benchmarks.
\emph{Methodology.} We analyze nine benchmarks commonly used to evaluate the effectiveness of agentic coding tools. We identify the constructs that the benchmarks claim to measure, the metrics used to operationalize them, and potential confounders that interfere with the measurement.
\emph{Results.} Our findings show that benchmarks commonly operationalize broad concepts such as coding capability with only a small set of metrics. The reported attributes that might confound the measurements vary between benchmarks, indicating a lack of standardization.
\emph{Conclusions.} We conclude with recommendations to strengthen construct validity in agentic coding benchmarks and encourage further empirical work to better understand the effectiveness of agentic coding tools.
\end{abstract}


\section{Introduction}
A benchmark is a systematic evaluation framework for assigning a quantitative score to a tool or method under assessment~\cite{benchmark-def}. The goal of benchmarking is to enable reproducible comparisons of different approaches by measuring a given property. Critical realism is a philosophy of science that assumes reality exists independently of our beliefs, while recognizing that our knowledge of it is fallible. From this perspective, measurement depends on theoretical assumptions and value judgments~\cite{archer2013critical}. For example, measuring the quality of agent-generated code requires a theory of code quality. People with different values and goals may disagree on the quality of the same code, even if they agree on an overall theory of code quality. When value conflict is low and the theory underlying a measurement is widely accepted and well understood, we say that the measurement is ``direct.'' Direct measurements are assumed to have inherent validity. In contrast, if there is value conflict or a measure depends on theories in which we are less confident, then the measurement's validity requires explicit justification. Such properties are modeled as \emph{constructs} and require explicit operationalization.
The \emph{construct validity} of an operationalization requires several forms of evidence. \emph{Face validity} concerns whether an indicator plausibly represents the construct; \emph{content validity} concerns coverage of its dimensions. \emph{Convergent validity} concerns expected relationships among indicators of the same construct, while \emph{discriminant validity} concerns their distinction from indicators of other constructs. \emph{Predictive validity} concerns whether measurements predict theoretically related outcomes. In reflective measurement models, convergent and discriminant validity are assessed using multiple indicators, with shared variance used to estimate the latent construct~\cite{mono-method}.
Since agentic coding tools such as Claude Code and Codex were introduced in 2025, researchers have proposed a range of benchmarks to evaluate their effectiveness for software engineering tasks~\cite{benchmarking}. For example, after the introduction of \emph{\swebench} by Jimenez et al., their benchmarking approach was applied to an increasingly diverse set of repositories spanning multiple programming languages~\cite{multi-swe}, human-verified task samples~\cite{openai-swebench-verified,deepswe}, or long-horizon tasks~\cite{swebench-pro}. Based on these benchmarks, companies such as OpenAI and Anthropic compare a newly introduced model, which can serve as the backbone of an agentic coding tool, against results from earlier model generations to demonstrate the superiority of the new model at the time of its release~\cite{google2026gemini4argon,openai2026gpt6astra,anthropic2026claudeopus55}. Similarly, SE researchers commonly adopt these benchmarks to compare approaches for improving agentic coding systems~\cite{agent-engineering,multi-agent,unified-se,agent-swe}.
However, despite the widespread adoption of agentic coding tools, recent work has raised various concerns about the validity of the estimated scores in current benchmarks~\cite{position-construct}. Understanding these threats can guide the design of new benchmarks whose task samples and evaluation metrics better represent the capabilities they claim to measure. Previous analyses have indicated threats to the validity of the benchmarks, arising from issues such as misalignment between the description of an issue and the associated code change~\cite{issue-mismatch}, weaknesses in the validation of a generated candidate solution~\cite{swe-solved-correctly}, or the contamination of benchmarks in the training of LLMs~\cite{swebench-illusion}.
We complement this prior work by presenting a preliminary study of potential threats to the construct validity of current agentic coding benchmarks. Our research questions are:
\begin{researchquestions}
\textbf{RQ1}\hspace{0.3em}\emph{What constructs do current benchmarks claim to measure and how are these constructs operationalized?}
\textbf{RQ2}\hspace{0.3em}\emph{What additional attributes are recorded for the tasks included in current agentic coding benchmarks?}
\end{researchquestions}

\section{Dataset}
We extracted coding benchmarks from model cards and evaluation reports for \opus~\cite{anthropic2026claudeopus55}, \astra~\cite{openai2026gpt6astra}, and \gemini~\cite{google2026gemini4argon}, the most recent frontier models at the time of our study (October 2026). \Cref{tab:overview} lists the nine candidate benchmarks and the reports evaluating each.
We summarize their tasks and evaluation procedures below.

\begin{table}[t]
    \centering
    \caption{Benchmarks included in this study. A checkmark indicates an evaluation in the model's report. We omit benchmark version numbers in the remainder of the paper. Model columns: G = \astra, O = \opus, Ge = \gemini.}
    \label{tab:overview}
    \footnotesize
    \setlength{\tabcolsep}{2pt}
    \begin{tabularx}{\columnwidth}{@{}>{\raggedright\arraybackslash}X r l c c c@{}}
    \toprule
        Benchmark & Year & Venue & G & O & Ge \\
        \midrule
        \emph{DeepSWE v1.1}~\cite{deepswe}
            & 2026 & arXiv
            & \ding{51} & \ding{51} & \ding{51} \\
        \emph{FrontierCode  v1.1 extended}~\cite{lu2026frontiercode}
            & 2026 & Blog post
            & \ding{51} & \ding{51} & \\
        \emph{FrontierSWE v2}~\cite{proximal2026frontierswev2}
            & 2026 & Blog post
            &  & \ding{51} & \ding{51} \\
        \emph{SWE-Atlas-QnA-AA}~\cite{swe-atlas}
            & 2026 & DL4C@ICML
            & \ding{51} & & \\
        \emph{\swebench Multilingual}~\cite{swe-bench-multilingual}
            & 2025 & NeurIPS
            & & \ding{51} & \\
        \emph{\swebench Multimodal}~\cite{swe-bench-multimodal}
            & 2025 & ICLR
            & & \ding{51} & \\
        \emph{\swebench Pro}~\cite{swebench-pro}
            & 2026 & ICML
            & & \ding{51} & \\
        \emph{Terminal-Bench 4.0}~\cite{terminal-bench,terminal-bench-v4}
            & 2026 & ICLR
            & \ding{51} & \ding{51} & \ding{51} \\
        \emph{Vibe Code Bench}~\cite{vibe-code-bench}
            & 2026 & CAIS
            & & & \ding{51} \\
    \bottomrule
    \end{tabularx}
\end{table}

\textbf{\emph{DeepSWE}} contains 113 tasks spanning 91 active open-source repositories across five programming languages. Tasks are written from scratch by humans rather than adapted from existing commits or pull requests, and hand-written behavior tests assess observable software behavior~\cite{deepswe}.
\textbf{\emph{FrontierCode}} comprises 150 tasks drawn from 36 open-source repositories and authored by their maintainers. Each task provides a natural-language specification and is evaluated against maintainer-defined criteria. Evaluation combines automated tests and other programmatic checks with assessment by an LLM judge~\cite{lu2026frontiercode}.
\textbf{\emph{FrontierSWE}} contains 34 tasks across five categories: implementation, scientific computing, performance optimization, visual reasoning, and AI research. Tasks run in executable environments and are evaluated using deterministic automated verifiers~\cite{proximal2026frontierswev2}.
\textbf{\emph{SWE-Atlas}} comprises 284 expert-authored tasks from 18 actively maintained open-source repositories: Codebase Q\&A (124 tasks), Test Writing (90), and Refactoring (70). Tasks use natural-language specifications, including underspecified and runtime-oriented scenarios, and are independently reviewed by three experts. Evaluation combines category-specific programmatic checks with rubrics assessing completeness, maintainability, and code quality~\cite{swe-atlas}.
\textbf{\emph{SWE-bench Multilingual}} contains 300 tasks derived from real-world GitHub issues and their resolving pull requests across 42 repositories and nine programming languages. Validation tests check issue resolution and preservation of existing functionality~\cite{swe-bench-multilingual}.
The original \textbf{\emph{SWE-bench Multimodal}} release contains 619 tasks from 17 JavaScript libraries~\cite{swe-bench-multimodal-data}. It covers interfaces, diagramming, data visualization, syntax highlighting, and interactive mapping. Tasks include visual content, such as images or videos, in their problem descriptions or testing scenarios. Candidate solutions are verified using tests~\cite{swe-bench-multimodal}.

\textbf{\emph{SWE-bench Pro}} contains \num{1865} tasks from 41 repositories, including 11 public, 12 held-out, and 18 proprietary repositories. Tasks are drawn from software projects and are divided into public, held-out, and commercial evaluation sets. Verification relies on automated test suites, including tests designed to check issue resolution and regression tests to detect unintended changes to existing functionality~\cite{swebench-pro}.

\textbf{\emph{Terminal-Bench}} comprises 66 tasks that evaluate agents' ability to complete complex tasks in containerized terminal environments~\cite{terminal-bench-v4}. The original \emph{Terminal-Bench} paper describes the benchmark's CLI-task framing and resolution-rate evaluation procedure~\cite{terminal-bench}.
\textbf{\emph{Vibe Code Bench}} comprises 100 natural-language web application specifications, divided equally into validation and held-out test sets, covering individual applications (24), solo-founder products (45), and enterprise tools (31). A GPT-5 pipeline generated the specifications, which were refined through two-pass human review and a final audit; applications are evaluated using predefined browser-based workflows, totaling 964 workflows and \num{10131} substeps~\cite{vibe-code-bench}.

\section{Methodology}
Through a qualitative analysis of the introductory benchmark publications, we extracted the intended constructs and the metrics used to operationalize them. Furthermore, we extracted additional task attributes that were used to identify potential construct-irrelevant variance.

The ACM SIGSOFT Empirical Standards require a new benchmark to specify four components: i) the quality to be benchmarked, ii) the metrics used to quantify it, iii) the measurement procedure, and iv) the workload or task sample~\cite{empirical-standards}.

\emph{\textbf{RQ1}}
For this research question, we provide an overview of \emph{what} the current benchmarks aim to measure and \emph{how} they measure it. To achieve this, we performed a qualitative analysis to identify the constructs communicated in the original literature that introduced the benchmarks. For our analysis, we combined human annotation with LLM-based analysis. That is, one of the authors read the papers and extracted the constructs that the benchmarks claim to measure. Since not all papers communicated an explicit construct that the benchmark aims to measure, the author inferred the related construct from the presentation of the curated task sample. For example, while \emph{\swebench Pro} does not explicitly communicate a construct to be measured, we inferred the target label \enquote{Software development capability} from the description of its task sample, stating that the benchmark \enquote{\textit{\elide more faithfully captures the complexity and diversity of real-world software development}}~\cite{swebench-pro}. We report the quotes that we used to label each benchmark in our supplementary material~\cite{zenodo}. The labels in \Cref{tab:metrics} summarize these extracted or inferred constructs, with \enquote{capability} added consistently to identify the intended ability. Inferred labels describe our interpretation of the task domain; they do not establish that the benchmark authors defined or validated the corresponding construct. Inferred labels are marked explicitly in \Cref{tab:metrics}.

\emph{\textbf{RQ2}}
To address our second research question, we analyze the extent to which benchmarks report task attributes beyond their outcome metrics. Such attributes enable stratified analyses of benchmark scores and the identification of potential sources of variance. For example, OpenAI reported model performance stratified by task difficulty on \emph{SWE-bench Verified}~\cite{openai-swebench-verified}. Similarly, Vergopoulos et al.~\cite{issue-quality} reported a negative correlation between fix complexity and agent performance.

\section{Results \& Reflections}
\label{sec:results}

\subsection{\textbf{RQ1}: Constructs and operationalization}
We report the constructs and main performance metrics per benchmark in \Cref{tab:metrics}. We excluded \emph{FrontierSWE} from \textbf{RQ1} because its introductory blog did not contain sufficient information for our analysis. We found explicit statements introducing the constructs of the benchmarks for \emph{DeepSWE} and \emph{\swebench Multilingual} and inferred the constructs for the others. Seven of the eight remaining benchmarks report task pass rates: the proportion of candidate solutions meeting the benchmark's acceptance criteria.
\begin{table}[t]
\centering
    \caption{Target labels and metrics. \textsuperscript{*} marks inferred labels. Labels denote intended domains; validity requires separate evidence.}
\label{tab:metrics}
\footnotesize
\setlength{\tabcolsep}{2pt}
\setlength{\aboverulesep}{1pt}
\setlength{\belowrulesep}{1pt}
\newcommand{\metric}[2]{\noindent(#1)\,#2\par}
\begin{tabularx}{\columnwidth}{@{}>{\scriptsize\raggedright\arraybackslash}p{85pt} >{\raggedright\arraybackslash}p{74pt} >{\raggedright\arraybackslash}X@{}}
\toprule
Benchmark & Target label & Metrics \\
\midrule
\emph{DeepSWE}~\cite{deepswe} & SE capability & \metric{1}{Pass rate} \\
\emph{FrontierCode}~\cite{lu2026frontiercode} & Coding capability\textsuperscript{*} & \metric{1}{Pass rate}\metric{2}{Weighted rubric score} \\
\emph{SWE-Atlas}~\cite{swe-atlas} & SE capability\textsuperscript{*} & \metric{1}{Pass rate}\metric{2}{Rubric-set pass rate} \\
\emph{SWE-bench Multilingual}~\cite{swe-bench-multilingual} & SE capability & \metric{1}{Pass rate} \\
\emph{SWE-bench Multimodal}~\cite{swe-bench-multimodal} & Bug fixing capability\textsuperscript{*} & \metric{1}{Pass rate} \\
\emph{SWE-bench Pro}~\cite{swebench-pro} & Softw. dev. capability\textsuperscript{*} & \metric{1}{Pass rate} \\
\emph{Terminal-Bench}~\cite{terminal-bench-v4} & CLI use capability\textsuperscript{*} & \metric{1}{Pass rate} \\
\emph{Vibe Code Bench}~\cite{vibe-code-bench} & Web dev. capability\textsuperscript{*} & \metric{1}{Workflow accuracy} \\
\bottomrule
\end{tabularx}
\end{table}
Rubric evaluations assess solutions against explicit criteria, with LLM judges used for properties that are difficult to check automatically. \emph{FrontierCode} additionally reports a weighted score over rubric items covering, for example, test quality, scope discipline, and codebase conventions; failure on any blocking criterion sets the score to zero~\cite{lu2026frontiercode}.
\emph{SWE-Atlas} uses binary rubric judgments for aspects such as answer completeness, test comprehensiveness, and refactoring maintainability. Task acceptance requires all mandatory rubric items and any required programmatic checks to pass; some nonmandatory items provide qualitative feedback. It also reports Test Writing and Refactoring rubric-set pass rates~\cite{swe-atlas}.
\emph{Vibe Code Bench} reports workflow accuracy: a workflow passes when at least 90\% of its substeps succeed according to the browser-agent evaluator. Application accuracy is the percentage of workflows that pass, and the benchmark score is the mean of these percentages across test applications~\cite{vibe-code-bench}.
The label \enquote{software engineering capability} is recorded for \emph{DeepSWE}, \emph{SWE-Atlas}, and \emph{SWE-bench Multilingual} in \Cref{tab:metrics}; it is inferred for \emph{SWE-Atlas} and interpreted within each benchmark's stated scope. \emph{DeepSWE} specifically targets long-horizon repository implementation, requiring larger code changes with comparatively shorter descriptions~\cite{deepswe}. \emph{FrontierCode}'s \enquote{coding capability} label concerns \enquote{\textit{mergeable code}}, i.e., code accepted for merging by a human reviewer~\cite{lu2026frontiercode}. We label \emph{SWE-bench Pro}'s construct \enquote{software development capability}, while the remaining labels specify a task type (\enquote{bug fixing capability}), a type of software (\enquote{web development capability}), or tool use (\enquote{CLI use capability}).
\emph{SWE-bench Multilingual}'s target concerns issue resolution across programming languages~\cite{swe-bench-multilingual}. Our inferred label for \emph{SWE-Atlas} covers codebase question answering, test writing, and refactoring~\cite{swe-atlas}. For \emph{SWE-bench Pro}, our inferred \enquote{software development capability} label refers specifically to producing repository changes from issue descriptions, verified through automated tests~\cite{swebench-pro}. These scopes do not establish capability across the full software engineering lifecycle.

\textbf{Reflection:} Our empirical findings raise several concerns regarding the construct validity of current software engineering benchmarks.
Failing to specify a target construct is a substantive measurement problem: readers cannot judge whether the selected tasks and evaluation criteria adequately represent the intended capability. Inferring a domain label from the task sample does not resolve this problem.
Interpreting an aggregate pass rate as a measure of broad coding or software engineering capability requires evidence that the tasks and evaluation criteria cover the intended construct. Missing dimensions would constitute construct underrepresentation~\cite{contruct-sjoberg,underrepresentation}. Prior measurement work recommends defining constructs and validating the relationships among multiple indicators, including their factor structure~\cite{arif-construct,mono-method}. This coverage question concerns content validity. The number of reported aggregate metrics alone does not establish whether the underlying tasks provide sufficient coverage.
Recommended instrumentation practice calls for defining related constructs and assessing measurement models with multiple indicators~\cite{mono-method}. Evidence on factor structure, predictive relationships, and score stability would support broader capability interpretations. We recorded target claims, aggregate metrics, and task attributes. We did not systematically audit validation procedures or estimate measurement models; their omission does not imply their absence.

These findings identify potential validity threats without establishing that any benchmark is invalid.
The rubric evaluations in \emph{FrontierCode} and \emph{SWE-Atlas} represent LLM judgments based on predefined criteria. In earlier SE research, Dieste et al. measured programmer performance using external functional code quality and productivity~\cite{performance-quality}. They operationalized productivity as the percentage of acceptance-test assertions passed during a fixed two-hour programming task. This proxy represents delivered functionality relative to a particular test suite and time budget; its suitability as a measure of broader programmer productivity requires additional validity evidence. The relevance of those measures to broader coding capability, and of rubric scores to nonfunctional quality, requires validation in future work.
However, selecting the right set of metrics to measure code quality remains an area of research~\cite{arif-construct, code-quality}, and producing high-quality code might not capture all aspects that the claimed constructs include. For example, a benchmark that primarily evaluates whether an agent can produce high-quality code that passes a set of tests may provide evidence of its implementation capability, but it does not necessarily establish its ability to understand requirements, design software architectures, review code, or maintain existing systems.
These activities represent different aspects of software development that may not be inferable from the verification of implemented candidate solutions given a task description. Consequently, even if we ignored the threats to the validity of benchmark scores raised in prior studies~\cite{issue-mismatch,swe-solved-correctly}, the scores should be interpreted in relation to the specific capabilities exercised by their tasks rather than as comprehensive measures of general software engineering capability.

\subsection{\textbf{RQ2}: Additional Aspects}
\begin{figure}[t]
    \centering
    \begin{tikzpicture}
    \begin{axis}[
        xbar,
        width=6.5cm,
        height=4.8cm,
        xlabel={Count},
        ylabel={Aspect},
        symbolic y coords={
            repository,
            {task category},
            language,
            difficulty,
            {task specificity},
        },
        ytick=data,
        font=\footnotesize,
        tick label style={font=\footnotesize},
        label style={font=\footnotesize},
        xmin=0,
        xmax=6,
        xtick={0,1,2,3,4,5,6},
        grid=both,
        axis lines=left,
        axis line style={-},
        tick style={black},
        enlarge y limits=0.15,
        enlarge x limits={upper,value=0.02},
    ]
    \addplot[
        fill=blue!55,
        draw=blue!75
    ] coordinates {
        (6,repository)
        (5,{task category})
        (4,language)
        (2,difficulty)
        (1,{task specificity})
    };
    \end{axis}
    \end{tikzpicture}
    \caption{Number of benchmarks reporting each additional task attribute. A benchmark can report multiple attributes.}
    \label{fig:aspects}
\end{figure}
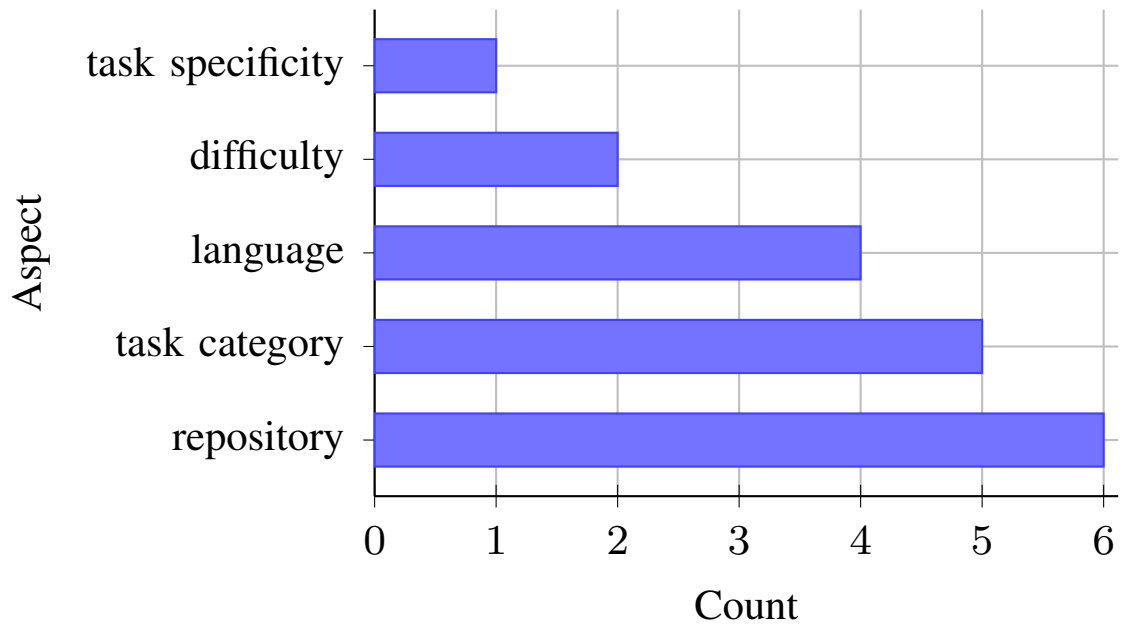

We report the additional aspects and the number of benchmarks in which they were captured in \Cref{fig:aspects}. In addition to \emph{FrontierSWE}, we excluded \emph{FrontierCode} from \textbf{RQ2} because its dataset is not publicly available, leaving seven benchmarks.
The most commonly reported task attribute is the repository name, appearing in six of seven benchmarks. \emph{Terminal-Bench} does not report a source repository uniformly because its tasks extend beyond coding in software projects. The next most commonly reported attribute is task category. The categories vary between datasets. \emph{SWE-Atlas} uses a hierarchy that first classifies tasks as \enquote{\textit{Codebase QnA}}, \enquote{\textit{Refactoring}}, or \enquote{\textit{Test Writing}}, and then assigns each task a subcategory within its category, e.g., \enquote{\textit{Root-cause analysis}} or \enquote{\textit{Unit Tests}}~\cite{swe-atlas}. In contrast, \emph{SWE-bench Pro} uses knowledge-area labels such as \enquote{\textit{database\_knowledge}} and \enquote{\textit{authentication\_authorization\_knowledge}}~\cite{swebench-pro}.
The programming language in which a task has to be implemented was explicitly reported per example in four of seven benchmarks. For the others, it would have to be inferred from the solution patches or the targeted repository. Task difficulty was reported in two of seven benchmarks. \emph{Terminal-Bench} reports an estimated expert completion time per task~\cite{terminal-bench-v4}. \emph{SWE-bench Pro}'s metadata includes a binary flag identifying hard tasks~\cite{swebench-pro-v2}. \emph{SWE-bench Pro} also reports task-specificity labels that distinguish, for example, enhancements from bug fixes. These serve a similar purpose to \emph{SWE-Atlas}'s task hierarchy without explicitly relating the labels.



\textbf{Reflection:}
Our findings show that benchmarks report potentially relevant task attributes. Previous studies reported performance differences across task categories~\cite{swe-atlas} and negative correlations between fix complexity and agent performance~\cite{issue-quality}. For agents that configure test environments, success rates also varied by programming language~\cite{pl}. In industrial C++ test generation, Bekmyradov et al.~\cite{vekil} reported differences between systems without attributing them to any single factor. Repository size, measured by LOC and file count, was likewise negatively associated with resolution rates~\cite{swe-bench-live}.
Aside from construct validity, we argue that our analysis reveals a lack of standardization across benchmarks. This inconsistency undermines comparability between benchmarks. We argue that differences in how attributes such as task category or task difficulty are reported across benchmarks encourage future empirical work to identify and validate the aspects or metrics that influence an agent's success rate. Given the large number of existing papers on benchmarking agentic coding tools~\cite{benchmarking,benchmarking2}, we also encourage secondary work to synthesize existing findings and provide an overview of the factors to consider.   

\section{Future Plans}
We divide our future plans into \emph{benchmarking} and \emph{validation}. Although our analysis identifies potential threats to the construct validity of existing benchmarks, the goal of our study is to encourage discussion of these threats rather than discourage future work on benchmarking agentic coding tools.

\emph{Benchmarking}
Our plans for future work on benchmark curation include extending existing benchmarks with additional code-quality metrics. We view the rubric-based approaches in \emph{FrontierCode} and \emph{SWE-Atlas} as a good starting point, and plan to investigate whether static analyses can extend or, where justified, replace assessments by LLM judges. Because Arif et al.'s validation evidence concerns object-oriented, class-level Java metrics~\cite{arif-construct}, applying those metrics to agent-generated changes across other languages and task types would require establishing which constructs they reflect and validating them in each target setting. Their use in agentic coding benchmarks remains an empirical research question. Our vision for agentic coding benchmarks is that they should provide insight into the long-term effects of using an agentic coding tool to generate changes in a codebase. Static analyses could examine code properties relevant to long-term maintenance, but would not by themselves establish the long-term effects of agent use. Established standards, such as the ISO standard for systems and software quality~\cite{iso25010}, may help identify the aspects a benchmark should cover.

\emph{Validation}
As the capabilities of recent LLMs have broadened, benchmarking them has become more complex than benchmarking traditional machine learning models or earlier language models. Because of this increasing complexity, we plan to empirically assess the factor structure underlying constructs related to the coding abilities of agents. Building on Khatib et al.'s study of how including information in task descriptions affects agent success~\cite{description-ablation}, we plan to examine more task categories and dimensions of description quality.

\clearpage

{\balance
\bibliographystyle{IEEEtran}
\bibliography{bibliography}

@online{openai-swebench-verified,
  author = {OpenAI},
  title  = {{Introducing SWE-bench Verified}},
  year   = {2024}
}

@inproceedings{benchmark-def,
  author       = {Inioluwa Deborah Raji and
                  Emily Denton and
                  Emily M. Bender and
                  Alex Hanna and
                  Amandalynne Paullada},
  editor       = {Joaquin Vanschoren and
                  Sai{-}Kit Yeung},
  title        = {{AI} and the Everything in the Whole Wide World Benchmark},
  booktitle    = {Proceedings of the Neural Information Processing Systems Track on
                  Datasets and Benchmarks 1, NeurIPS Datasets and Benchmarks 2021, December
                  2021, virtual},
  year         = {2021},
  url          = {https://datasets-benchmarks-proceedings.neurips.cc/paper/2021/hash/084b6fbb10729ed4da8c3d3f5a3ae7c9-Abstract-round2.html},
  bibsource    = {dblp computer science bibliography, https://dblp.org}
}

@inproceedings{issue-mismatch,
  author    = {Manyi Wang and Junjielong Xu and Pinjia He},
  title     = {PAIChecker: Uncovering and Checking PR-Issue Misalignment in SWE-Bench-Like Benchmarks},
  booktitle = {ASE},
  year      = {2026}
}

@inproceedings{terminal-bench,
  author    = {Mike A. Merrill et al.},
  title     = {Terminal-Bench: Benchmarking Agents on Hard, Realistic Tasks in Command Line Interfaces},
  booktitle = {ICLR},
  year      = {2026}
}

@misc{zenodo,
  author       = {Author, Anonymous},
  title        = {{Construct Validity of Coding Benchmarks for LLM-based Agentic Coding Tools}},
  month        = oct,
  year         = 2026,
  publisher    = {Zenodo},
  howpublished = {Zenodo},
  doi          = {10.5281/zenodo.23599441},
  url          = {https://doi.org/10.5281/zenodo.23599441}
}

@inproceedings{swebench-illusion,
  author    = {Shanchao Liang and Spandan Garg and Roshanak Zilouchian Moghaddam},
  title     = {The SWE-Bench Illusion: When State-of-the-Art LLMs Remember Instead of Reason},
  booktitle = {ICSE-SEIP},
  year      = {2026}
}

@inproceedings{unified-se,
  author    = {Applis, Leonhard and Zhang, Yuntong and Liang, Shanchao and Jiang, Nan and Tan, Lin and Roychoudhury, Abhik},
  title     = {Unified Software Engineering Agent as AI Software Engineer},
  booktitle = {ICSE},
  year      = {2026}
}

@inproceedings{multi-agent,
  author    = {Li, Han and Shi, Yuling and Lin, Shaoxin and Gu, Xiaodong and Lian, Heng and Wang, Xin and Jia, Yantao and Huang, Tao and Wang, Qianxiang},
  title     = {SWE-Debate: Competitive Multi-Agent Debate for Software Issue Resolution},
  booktitle = {ICSE},
  year      = {2026}
}

@inproceedings{agent-engineering,
  author    = {Tian, Zhao and Gao, Pengfei and Chen, Junjie and Peng, Chao},
  title     = {Agent-Based Ensemble Reasoning for Repository-Level Issue Resolution},
  booktitle = {ICSE},
  year      = {2026}
}

@inproceedings{swe-bench-multilingual,
  author    = {John Yang and Kilian Lieret and Carlos E. Jimenez and Alexander Wettig and Kabir Khandpur and Yanzhe Zhang and Binyuan Hui and Ofir Press and Ludwig Schmidt and Diyi Yang},
  title     = {SWE-smith: Scaling Data for Software Engineering Agents},
  booktitle = {NeurIPS},
  year      = {2025}
}

@online{lu2026frontiercode,
  author       = {Lu, E. and Pan, B. and Birlikci, D. and Lee, S. and Wang, R. and Choudhury, R. and Ma, F. and Qin, T. C. and Baronio, C. and Alberti, S. and others},
  title        = {Introducing FrontierCode},
  organization = {Cognition},
  month        = jun,
  year         = {2026}
}

@techreport{google2026gemini4argon,
  author      = {{Google DeepMind}},
  title       = {Gemini 4 Argon Model Card},
  institution = {Google DeepMind},
  type        = {Model Card},
  month       = oct,
  year        = {2026}
}

@online{openai2026gpt6astra,
  author = {{OpenAI}},
  title  = {{GPT-6 Astra}: A new generation of intelligence},
  month  = sep,
  year   = {2026}
}

@techreport{anthropic2026claudeopus55,
  author      = {{Anthropic}},
  title       = {Claude Opus 5.5 System Card},
  institution = {Anthropic},
  type        = {System Card},
  month       = sep,
  year        = {2026}
}

@inproceedings{arif-construct,
  author    = {Hera Arif and Miikka Kuutila and Paul Ralph},
  title     = {Assessing the Construct Validity of Object-Oriented, Class-Level Code Quality Metrics},
  booktitle = {ESEM},
  year      = {2026}
}

@article{performance-quality,
  author  = {Oscar Dieste and Alejandrina M. Aranda and Fernando Uyaguari Uyaguari and Burak Turhan and Ayse Tosun and Davide Fucci and Markku Oivo and Natalia Juristo},
  title   = {Empirical evaluation of the effects of experience on code quality and programmer productivity: an exploratory study},
  journal = {Empir. Softw. Eng.},
  volume  = {22},
  number  = {5},
  year    = {2017}
}

@article{contruct-sjoberg,
  author  = {Dag I. K. Sj{\o}berg and Gunnar Rye Bergersen},
  title   = {Construct Validity in Software Engineering},
  journal = {{IEEE} Trans. Software Eng.},
  volume  = {49},
  number  = {3},
  year    = {2023}
}

@inproceedings{issue-quality,
  author    = {Konstantinos Vergopoulos and Mark Niklas M{\"{u}}ller and Martin T. Vechev},
  title     = {Automated Benchmark Generation for Repository-Level Coding Tasks},
  booktitle = {ICML},
  year      = {2025}
}

@inproceedings{vekil,
  author    = {Vekil Bekmyradov and Thomas Bach and Alexander Berndt and Noah Christoph P{\"{u}}tz and Bartosz Bogacz and Thomas Bartz{-}Beielstein},
  title     = {Evaluating LLM-Based Test Generation for a Large Industrial {C++} Database System. {A} Case Study on {SAP} {HANA}},
  booktitle = {ESEM},
  year      = {2026}
}

@article{swe-solved-correctly,
  author  = {You Wang and Michael Pradel and Zhongxin Liu},
  title   = {Are "Solved Issues" in SWE-bench Really Solved Correctly? An Empirical Study},
  journal = {CoRR},
  volume  = {abs/2503.15223},
  year    = {2025}
}

@article{code-quality,
  author  = {Cristina Improta and Pietro Liguori and Domenico Cotroneo},
  title   = {What is the Difference Between Me and You? Benchmarking the Quality Gap Between Human-Written and AI-Generated Code},
  journal = {CoRR},
  volume  = {abs/2609.12708},
  year    = {2026}
}

@book{archer2013critical,
  title={Critical Realism: Essential Readings},
  author={Archer, M. and Bhaskar, R. and Collier, A. and Lawson, T. and Norrie, A.},
  isbn={9781136287183},
  series={Ontological Explorations (Routledge Critical Realism)},
  url={https://books.google.ca/books?id=-wz6N9uBpnYC},
  year={2013},
  publisher={Taylor \& Francis}
}

@incollection{mono-method,
  author    = {Paul Ralph and Miikka Kuutila and Hera Arif and Bimpe Ayoola},
  title     = {Teaching Software Metrology: The Science of Measurement for Software Engineering},
  booktitle = {Handbook on Teaching Empirical Software Engineering},
  publisher = {Springer},
  year      = {2024}
}

@article{underrepresentation,
author = {Messick, Samuel},
title = {VALIDITY},
journal = {ETS Research Report Series},
volume = {1987},
number = {2},
pages = {i-208},
doi = {https://doi.org/10.1002/j.2330-8516.1987.tb00244.x},
url = {https://onlinelibrary.wiley.com/doi/abs/10.1002/j.2330-8516.1987.tb00244.x},
eprint = {https://onlinelibrary.wiley.com/doi/pdf/10.1002/j.2330-8516.1987.tb00244.x},
year = {1987}
}

@inproceedings{agent-swe,
  author       = {Hasen {\"{O}}zayt{\"{u}}rk and
                  Feza Buzluca},
  editor       = {Robert Feldt and
                  Maria Paasivaara and
                  Daniel M{\'{e}}ndez and
                  Stefan Wagner and
                  Marvin Mu{\~{n}}oz Bar{\'{o}}n},
  title        = {Coordinating Agents on a Shared Git Workspace: An Empirical Study
                  of Action-Space Observation for Agentic Software Engineering},
  booktitle    = {20th International Symposium on Empirical Software Engineering and
                  Measurement, {ESEM} 2026, Munich, Germany, October 8-9, 2026},
  series       = {LIPIcs},
  volume       = {394},
  pages        = {18:1--18:20},
  publisher    = {Schloss Dagstuhl - Leibniz-Zentrum f{\"{u}}r Informatik},
  year         = {2026},
  url          = {https://doi.org/10.4230/LIPIcs.ESEM.2026.18},
  doi          = {10.4230/LIPICS.ESEM.2026.18},
  bibsource    = {dblp computer science bibliography, https://dblp.org}
}

@article{empirical-standards,
  author  = {Paul Ralph and Nauman bin Ali and Sebastian Baltes and Domenico Bianculli and Jessica Diaz and Yvonne Dittrich and Neil Ernst and Michael Felderer and Robert Feldt and Antonio Filieri and Breno Bernard Nicolau de França and Carlo Alberto Furia and Greg Gay and Nicolas Gold and Daniel Graziotin and Pinjia He and Rashina Hoda and Natalia Juristo and Barbara Kitchenham and Valentina Lenarduzzi and Jorge Martínez and Jorge Melegati and Daniel Mendez and Tim Menzies and Jefferson Molleri and Dietmar Pfahl and Romain Robbes and Daniel Russo and Nyyti Saarimäki and Federica Sarro and Davide Taibi and Janet Siegmund and Diomidis Spinellis and Miroslaw Staron and Klaas Stol and Margaret-Anne Storey and Damian Tamburri and Marco Torchiano and Christoph Treude and Burak Turhan and Xiaofeng Wang and Sira Vegas},
  title   = {Empirical Standards for Software Engineering Research},
  journal = {CoRR},
  volume  = {abs/2010.03525},
  year    = {2020}
}

@inproceedings{swebench-pro,
  author    = {Xiang Deng and Jeff Da and Edwin Pan and Yannis Yiming He and Charles Ide and Kanak Garg and Niklas Lauffer and Andrew Park and Chetan Rane and Karmini Sampath and Maya Krishnan and Srivatsa R Kundurthy and Sean M. Hendryx and Zifan Wang and Chen Bo Calvin Zhang and Noah Jacobson and Bing Liu and Brad Kenstler},
  title     = {{SWE}-Bench Pro: Can {AI} Agents Solve Long-Horizon Software Engineering Tasks?},
  booktitle = {ICML},
  year      = {2026}
}

@article{deepswe,
  author  = {Wenqi Huang and Charley Lee and Leonard Tng and Serena Ge},
  title   = {DeepSWE: Measuring Frontier Coding Agents on Original, Long-Horizon Engineering Tasks},
  journal = {CoRR},
  volume  = {abs/2607.07946},
  year    = {2026}
}

@inproceedings{benchmarking,
  author    = {Mahmoud Mohammadi and Yipeng Li and Jane Lo and Wendy Yip},
  title     = {Evaluation and Benchmarking of {LLM} Agents: {A} Survey},
  booktitle = {KDD},
  year      = {2025}
}

@article{swe-atlas,
  author  = {Mohit Raghavendra and Soham Dan and Miguel Romero Calvo and Yannis Yiming He and Johannes Baptist Mols and Gautam Anand and Cole McCollum and Edgar Arakelyan and Vijay Bharadwaj and Andrew Park and Jeff Da and MohammadHossein Rezaei and Bing Liu and Brad Kenstler and Yunzhong He},
  title   = {{SWE} Atlas: Benchmarking Coding Agents Beyond Issue Resolution},
  journal = {CoRR},
  volume  = {abs/2605.08366},
  year    = {2026}
}

@inproceedings{vibe-code-bench,
  author    = {Hung Tran and Langston Nashold and Rayan Krishnan and Antoine Bigeard and Alex Gu},
  title     = {Vibe Code Bench: Evaluating {AI} Models on End-to-End Web Application Development},
  booktitle = {CAIS},
  year      = {2026}
}

@inproceedings{swe-bench-multimodal,
  author    = {John Yang and Carlos E. Jimenez and Alex L. Zhang and Kilian Lieret and Joyce Yang and Xindi Wu and Ori Press and Niklas Muennighoff and Gabriel Synnaeve and Karthik R. Narasimhan and Diyi Yang and Sida Wang and Ofir Press},
  title     = {SWE-bench Multimodal: Do {AI} Systems Generalize to Visual Software Domains?},
  booktitle = {ICLR},
  year      = {2025}
}

@article{proximal2026frontierswev2,
  author  = {Rishyanth Kondra and Sanket Mhatre and Akshit Kumar and Evan Chu and Bilal Bakht Ahmad and Ayush Nangia and Rajan Agarwal and Arpan Dasgupta and Animesh Sinha and Bhuvanesh Sridharan and Krupa Dave and Brendan Graham and Guanyu Song and Anirudh Rahul and Wei Hern Lim and Abishek Thangamuthu and Ramneet Singh and Danna Liu and Navid Pour and Calvin Chen and Justus Mattern},
  title   = {FrontierSWE v2},
  journal = {Proximal Blog},
  year    = {2026}
}

@inproceedings{multi-swe,
  author    = {Daoguang Zan and Zhirong Huang and Wei Liu and Hanwu Chen and Shulin Xin and Linhao Zhang and Qi Liu and Aoyan Li and Lu Chen and Xiaojian Zhong and Siyao Liu and Yongsheng Xiao and Liangqiang Chen and Yuyu Zhang and Jing Su and Tianyu Liu and Rui Long and Ming Ding and Liang Xiang},
  title     = {Multi-SWE-bench: {A} Multilingual Benchmark for Issue Resolving},
  booktitle = {NeurIPS},
  year      = {2025}
}

@article{description-ablation,
  author  = {Lara Khatib and Noble Saji Mathews and Meiyappan Nagappan and Pengyu Nie and Thomas Zimmermann},
  title   = {What Makes a Good Bug Report for an {AI} Agent?},
  journal = {CoRR},
  volume  = {abs/2607.07593},
  year    = {2026}
}

@online{iso25010,
  title    = {{ISO/IEC 25010}},
  year    = {2022},
  url     = {https://iso25000.com/index.php/en/iso-25000-standards/iso-25010},
  note = {Accessed on 2026-10-23},
}

@inproceedings{benchmarking2,
  author    = {Asaf Yehudai and Lilach Eden and Alan Li and Guy Uziel and Yilun Zhao and Roy Bar{-}Haim and Arman Cohan and Michal Shmueli{-}Scheuer},
  title     = {A Survey on Evaluation of LLM-based Agents},
  booktitle = {Findings of ACL},
  year      = {2026}
}

@inproceedings{swe-bench-live,
  author    = {Linghao Zhang and Shilin He and Chaoyun Zhang and Yu Kang and Bowen Li and Chengxing Xie and Junhao Wang and Maoquan Wang and Yufan Huang and Shengyu Fu and Elsie Nallipogu and Qingwei Lin and Yingnong Dang and Saravan Rajmohan and Dongmei Zhang},
  title     = {SWE-bench Goes Live!},
  booktitle = {NeurIPS},
  year      = {2025}
}

@inproceedings{pl,
  author    = {Kelin Fu and Tianyu Liu and Zeyu Shang and Yingwei Ma and Jiaheng Liu and Jian Yang and Kaigui Bian},
  title     = {Multi-Docker-Eval: {A} 'Shovel of the Gold Rush' Benchmark on Automatic Environment Building for Software Engineering},
  booktitle = {Findings of ACL},
  year      = {2026}
}

@online{terminal-bench-v4,
  author = {Ryan Marten},
  title  = {{Terminal-Bench 4.0}},
  month  = aug,
  year   = {2026},
  note   = {Dataset version 4.0.0}
}

@online{swebench-pro-v2,
  author = {{Scale AI}},
  title  = {{SWE-bench Pro V2} dataset},
  year   = {2026},
  note   = {V2 configuration}
}

@misc{swe-bench-multimodal-data,
  author       = {{SWE-bench Team}},
  title        = {{SWE-bench Multimodal}: Original dataset release},
  howpublished = {Hugging Face dataset},
  year         = {2024},
  note         = {Revision 0ac4e9c0bc8f9cc717412e4851a67ef7b004d37c}
}

@article{position-construct,
  author       = {Maria I. Gorinova and
                  Macey Baker and
                  Amy Heineike and
                  Maksim Shaposhnikov and
                  Rob Willoughby and
                  Dru Knox},
  title        = {Position: Coding Benchmarks Are Misaligned with Agentic Software Engineering},
  journal      = {CoRR},
  volume       = {abs/2606.17799},
  year         = {2026},
  url          = {https://doi.org/10.48550/arXiv.2606.17799},
  doi          = {10.48550/ARXIV.2606.17799},
  eprinttype   = {arXiv},
  eprint       = {2606.17799},
  bibsource    = {dblp computer science bibliography, https://dblp.org}
}
}
\end{document}